# The study of amplitude and phase relaxation impact on the quality of quantum information technologies

Yu.I. Bogdanov[*a,b,c], B.I. Bantysh[†a,b], A.Yu. Chernyavskiy[a,d], V.F. Lukichev[a], A.A. Orlikovsky[a]

[a]Institute of Physics and Technology, Russian Academy of Science; [b]National Research University of Electronic Technology MIET; [c]National Research Nuclear University 'MEPHI'; [d]Faculty of Computational Mathematics and Cybernetics, Moscow State University

**ABSTRACT**

The influence of amplitude and phase relaxation on evolution of quantum states within the formalism of quantum operations is considered. The model of polarizing qubits where noises are determined by the existence of spectral degree of freedom that shows up during the light propagation inside anisotropic mediums with dispersion is studied. Approximate analytic model for calculation of phase plate impact on polarizing state with dispersion influence taken into consideration is suggested.

## 1. INTRODUCTION

One of the most significant bottlenecks in development of quantum informatics is an inevitable decoherence of quantum states during the implementation of quantum transformations. In this connection it becomes highly important to build a proper methodology for controlling of quantum states and processes. Therefore the purpose of this work is to study the influence of the environment on the quality of quantum gates and their properties.

The paper has the following structure. In the section 2 is given a brief introduction in the quantum operation theory, which includes such popular approaches of building non-ideal quantum gates models as the Kraus decomposition and Choi-Jamiolkowski relative states [1, 2]. It is shown in the section 3 that the usage of Choi-Jamiolkowski relative states helps us to give an account of the ability of quantum gates to produce the resource of quantum entanglement, which appears in every quantum algorithm, and the impact of quantum noises on this ability. The section 4 is devoted to the studying of quantum noise origination in the optical polarizing qubit. It is demonstrated there that the unitary transform that phase plate performs on polarizing state becomes non-unitary after taking radiation spectral degree of freedom into consideration. The final section 5 summarize the main conclusions of this work.

## 2. QUANTUM OPERATIONS SIMULATION WITH QUANTUM NOISES TAKEN INTO ACCOUNT

No real physical system is isolated from its surroundings and therefore it is constantly exchanging information with it. The interaction with the environment makes it possible to perform quantum state transformations. However, if these transformations contribute to the entanglement between the system under study and its environment, they lead to the loss of the system state coherence. If it is necessary to remain within the system's dimension, then such processes cannot be described with the language of state vectors $|\psi\rangle$ and unitary transformations $U$. In such cases, the quantum state is generally described by means of the density matrix $\rho$ and the Kraus operators $E_k$ are used for the description of quantum transformations:

---
[*] E-mail: bogdanov_yurii@inbox.ru
[†] E-mail: bbantysh60000@gmail.com

$$\rho(t) = \sum_k E_k(t)\rho_0 E_k^+(t). \tag{1}$$

The only condition imposed on the set of $E_k$ operators is the normalization condition, wherefrom

$$\sum_k E_k^+(t) E_k(t) = I. \tag{2}$$

In the case of unitary transformation $U$ there is a single element in the set of Kraus operators, which is equal to $U$. The interaction between the system and its environment can occur in absolutely diverse ways. In this work we focus on the processes of amplitude and phase relaxation, which were firstly studied during observation of spin-spin and spin-lattice relaxation of nuclear spins [3]. This processes are characterized by the time parameters $T_1$ и $T_2$ respectively and the following Kraus operators:

$$E_0^a = \begin{pmatrix} 1 & 0 \\ 0 & \sqrt{1-\gamma_a} \end{pmatrix}, \quad E_1^a = \begin{pmatrix} 0 & \sqrt{\gamma_a} \\ 0 & 0 \end{pmatrix} \quad \text{for amplitude relaxation,} \tag{3}$$

$$E_0^p = \begin{pmatrix} 1 & 0 \\ 0 & \sqrt{1-\gamma_p} \end{pmatrix}, \quad E_1^p = \begin{pmatrix} 0 & 0 \\ 0 & \sqrt{\gamma_p} \end{pmatrix} \quad \text{for phase relaxation,} \tag{4}$$

where

$$\sqrt{1-\gamma_a} = \exp\left(-\frac{t}{2T_1}\right), \quad \sqrt{1-\gamma_p} = \exp\left(-\frac{t}{2T_2^{pure}}\right), \quad T_2^{pure} = \frac{2T_1 T_2}{2T_1 - T_2}.$$

Along with the Kraus decomposition (1) there are other ways to describe quantum states evolution. Among them, the $\chi$-matrix formalism, or Choi-Jamiolkowski formalism [1, 2], can be distinguished. The approach implies a transition from an arbitrary set of Kraus operators to the fixed basis using the transition matrix $\chi$. In such case, any evolution of $s$-dimensional system can be represented as a matrix of dimension $s^2 \times s^2$. The quantum transformation $\chi$-matrix is determined in a unique way, however, it can correspond to different sets of Kraus operators. The $\chi$-matrix can be obtained by forming a maximally entangled superposition state between the system under study $A$ and an auxiliary system $B$ of the same dimension:

$$|\Phi\rangle = \frac{1}{\sqrt{s}} \sum_{j=1}^{s} |j\rangle \otimes |j\rangle. \tag{5}$$

It can be shown that in the result of the transformation $E$ to the subsystem $A$ we obtain the transformation $\chi$-matrix that is normalized to unity $\rho_\chi = \frac{1}{s}\chi$ (Fig. 1). It follows then that the $\chi$-matrix among other things is a density matrix in the system of a higher dimension. This statement forms the content of so-called Choi-Jamiolkowski isomorphism [4].

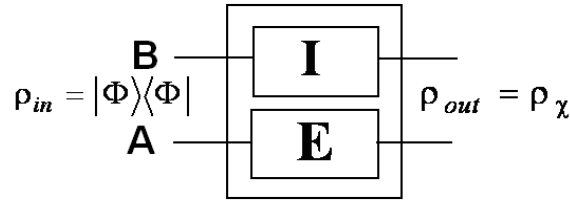

Fig. 1. Quantum scheme for $\chi$-matrix calculation

One of the main problems that we have to deal with during simulation of quantum operations with quantum noises is that noises are acting on the system simultaneously with the quantum gate. A solution to this problem may be derived under the assumption that Markov processes are taking place in such systems. In this case, we can divide the actual transformation $U$ into $N$ equal parts that act for time $\Delta t$ each to consider amplitude and phase relaxation processes in the periods. If the Choi-Jamiolkowski state is exposed to this complex transformation, then the resulting density matrix will correspond to the normalized $\chi$-matrix of noisy gate $U$. Figure 2 shows the example of the $\chi$-matrix of *SQiSW* transformation which is acting under the amplitude and phase relaxation with parameters $T_1 = 20T$ and $T_2 = 15T$, where $T$ is the gate execution time.

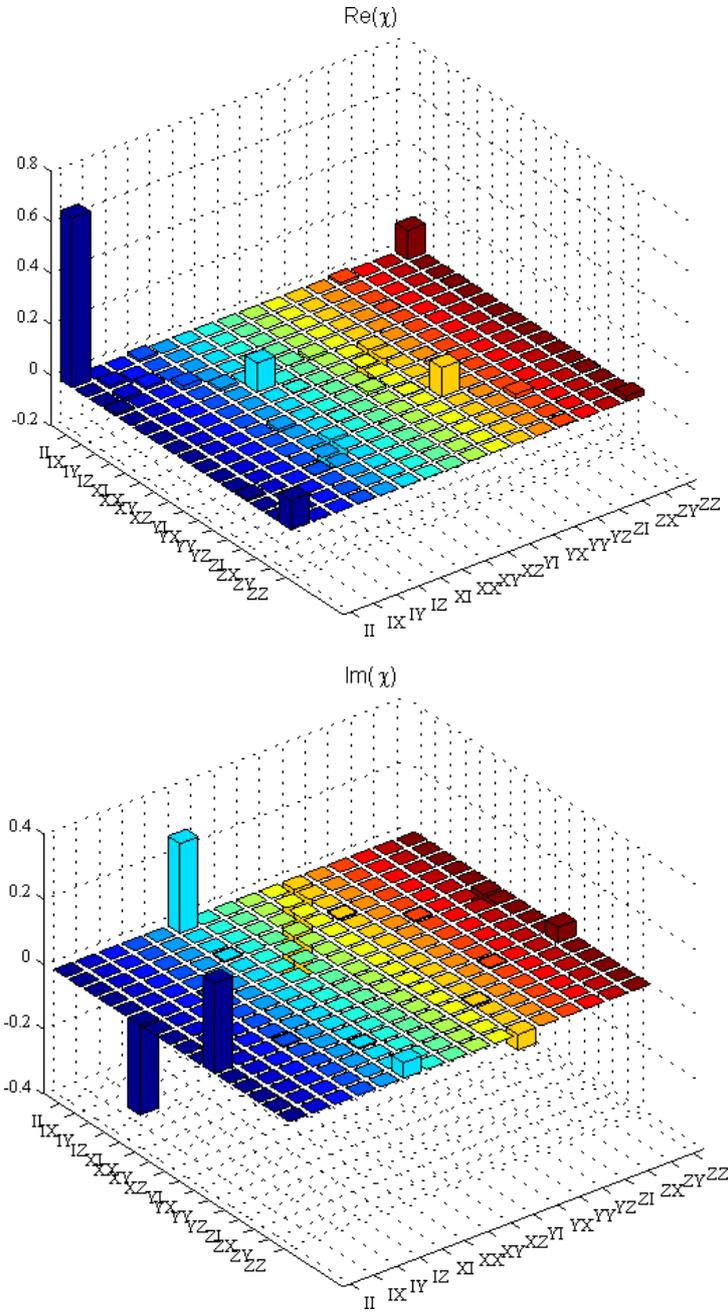

Fig. 2. $\chi$-matrix of *SQiSW* transformation under the influence of amplitude and phase relaxation with parameters $T_1 = 20T$ and $T_2 = 15T$

## 3. QUANTUM STATES ENTANGLEMENT UNDER THE INFLUENCE OF QUANTUM OPERATION

The fact that the χ-matrix of a quantum state, which is acting in the *s*-dimensional space, is also a density matrix in the $s^2$-dimensional space lets us to define a number of other characteristics of an arbitrary quantum transformation *E*. The analysis of changes in Choi-Jamiolkowski state properties under the influence of this transformation on subsystem *A* (Fig. 1) gives us an opportunity to see the effects on any quantum state that is subjected to *E* transformation [5].
One of such properties is the subsystems entanglement. To estimate the degree of entanglement we shall use the measure called Negativity, which is equal to the sum of all negative eigenvalues of a partially transposed density matrix:

$$Negativity = \frac{1}{2}\left(\sum_j |\lambda_j^{pt}| - \sum_j \lambda_j^{pt}\right), \tag{6}$$

where $\lambda_j^{pt}$ are eigenvalues of density matrix, transposed by one of the two subsystems. Figure 3 shows the result of simulation of negativity dynamics for a system under the influence of a *SQiSW* gate in the ideal case and in the case of amplitude and phase relaxation processes with parameters $T_1 = 20T$ and $T_2 = 15T$.

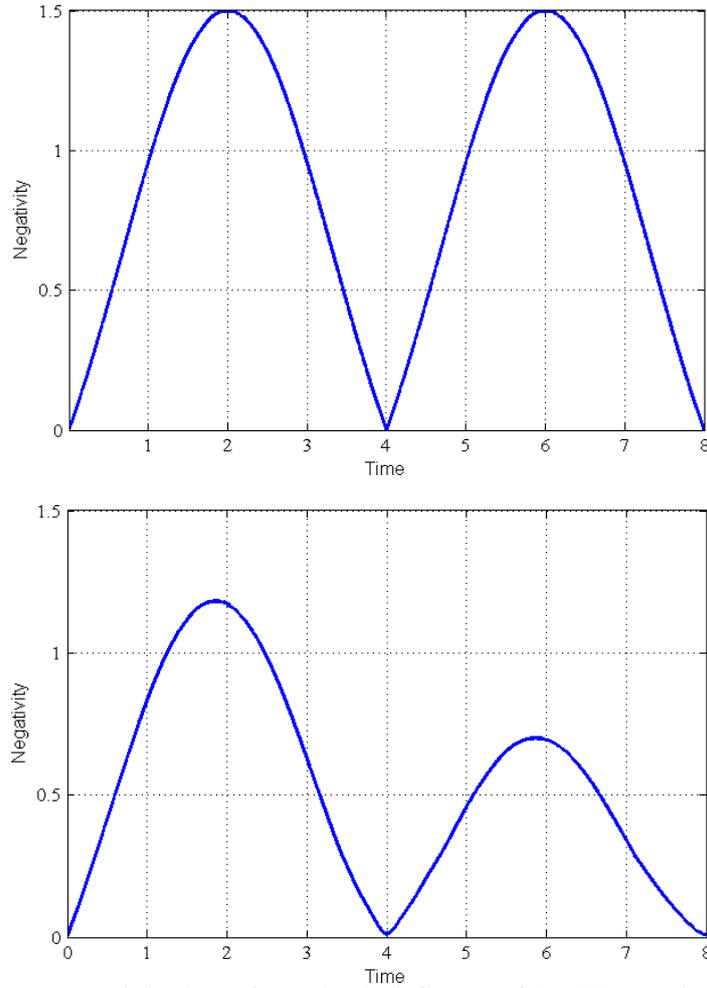

Fig. 3. Choi-Jamiolkowski state negativity dynamics under the influence of *SQiSW* operation in the ideal case and in the case of amplitude and phase relaxation processes with parameters $T_1 = 20T$ and $T_2 = 15T$

# 4. POLARIZING QUBIT PHASE RELAXATION

Depending on quantum bit realization, the processes which lead to amplitude and phase relaxation of a state can have a completely different nature, and it is not always possible to obtain simple analytical models to describe them. Let us consider a way to build a phase relaxation model for a polarizing qubit where the direction of light polarization is responsible for information encoding.

The essential element of any quantum optical scheme based on polarizing state is the phase plate, which transforms states with vertical |V> and horizontal |H> polarizations in the following way:

$$|V\rangle \rightarrow \left(\cos\left(\frac{\delta}{2}\right) - i\sin\left(\frac{\delta}{2}\right)\cos(2\alpha)\right)|V\rangle - i\sin\left(\frac{\delta}{2}\right)\sin(2\alpha)|H\rangle,$$
$$|H\rangle \rightarrow -i\sin\left(\frac{\delta}{2}\right)\sin(2\alpha)|V\rangle + \left(\cos\left(\frac{\delta}{2}\right) + i\sin\left(\frac{\delta}{2}\right)\cos(2\alpha)\right)|H\rangle, \quad (7)$$

where

$$\delta(\lambda) = \frac{2\pi \Delta n(\lambda) h}{\lambda}$$

is the optical length of the crystal, $\alpha$ is the angle between the optical axis of the crystal and the z-axis, $\lambda$ is the radiation wavelength, $h$ is the plate thickness and $\Delta n(\lambda) = |n_e(\lambda) - n_o(\lambda)|$ is the difference between the fast and slow axis refractive indexes. The transformation (7) is unitary, however the presence of the spectral degree of freedom in the polarizing state leads to its effective non-unitarity. Indeed, because the plate optical length depends on wavelength, different spectral components undergo different transformations. As a result, the original "pseudo-pure" state where all spectral components belonged to the same quantum state becomes mixed under the influence of the phase plate.

To obtain the $\chi$-matrix of the quantum operation performed by a phase plate, let us return to the notion that the normalized $\chi$-matrix is a density matrix in the space of a higher dimension. This lets us to write down the Choi-Jamiolkowski state in the output of the phase plate in a common formula that describes mixed states:

$$\rho_\chi = \sum_{\lambda_i} p(\lambda_i) \rho_{\lambda_i}. \quad (8)$$

The summation there is made by all the radiation spectral components. The natural thing to do there is to switch to integration, which can be solved using the linear Taylor expansion near the center $\lambda_0$ of the spectral distribution:

$$\delta(\lambda) = a + b(\lambda - \lambda_0), \quad a = \delta(\lambda_0), \quad b = \left(\frac{\partial \delta(\lambda)}{\partial \lambda}\right)_{\lambda = \lambda_0}. \quad (9)$$

The result of the simple calculations below is a $\chi$-matrix, written in an analytical form:

$$\chi = 2 \cdot \left(\rho_{11} |\varphi_1\rangle\langle\varphi_1| + \rho_{22} |\varphi_2\rangle\langle\varphi_2| + \rho_{12} |\varphi_1\rangle\langle\varphi_2| + \rho_{21} |\varphi_2\rangle\langle\varphi_1|\right), \quad (10)$$

where we introduce vectors

$$|\varphi_1\rangle = \frac{1}{\sqrt{2}} \begin{pmatrix} 1 \\ 0 \\ 0 \\ 1 \end{pmatrix}, \quad |\varphi_2\rangle = \frac{1}{\sqrt{2}} \begin{pmatrix} n_z \\ n_x \\ n_x \\ -n_z \end{pmatrix} \quad (11)$$

and the elements of some $2\times 2$ matrix

$$\rho = \frac{1}{2}\begin{pmatrix} 1+I_c\cos a - I_s\sin a & i(I_s\cos a + I_c\sin a) \\ -i(I_s\cos a + I_c\sin a) & 1 - I_c\cos a + I_s\sin a \end{pmatrix}. \tag{12}$$

$I_c$ and $I_s$ coefficients in the expression (12) represent the real and imaginary parts of continuous Fourier transform by radiation spectral distribution:

$$I_c = \int_{-\infty}^{\infty} \cos(b(\lambda-\lambda_0))p(\lambda)d\lambda, \quad I_s = \int_{-\infty}^{\infty} \sin(b(\lambda-\lambda_0))p(\lambda)d\lambda. \tag{13}$$

It can be easily seen that $I_s = 0$ for any symmetrical distribution. The described approach also reveals the fact that the rank of the phase plate $\chi$-matrix does not exceed 2 while being equal to 4 in the general case of a one-qubit transformation.

The consideration of $\chi$-matrix as a density matrix allows us to obtain the picture of the ability of a phase plate to preserve polarizing state purity depending on parameters of the plate and radiation. Such phase plate "purity" can be expressed with the following formula

$$P = \mathrm{Tr}(\rho^2) = \frac{1 + I_c^2 + I_s^2}{2}. \tag{14}$$

Figure 4 shows the dependence of phase plate "purity" on its thickness for various practically important spectral distributions.

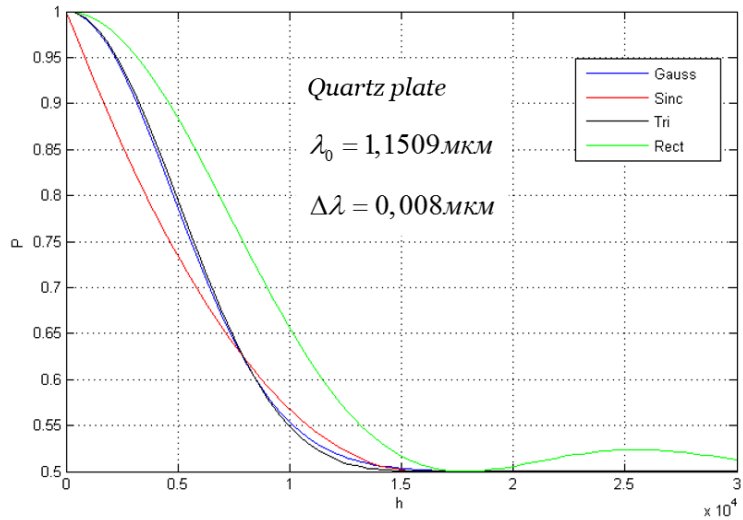

Fig. 4. The dependence of phase plate "purity" on its thickness for various radiation spectral distributions: Gaussian (Gauss), sinc$^2$ (Sinc), triangular (Tri) and uniform (Rect).

The linear approximation of plate optical length described above closely agrees with the results of mathematical simulations.
This study can be applied in the research of polarizing echo effects both analytically and experimentally [6].

## 5. CONCLUSION

Within the formalism of quantum operations, the influence of amplitude and phase relaxation on quantum states evolution is studied. The amplitude and phase relaxation impact on quantum correlations, which characterize the entanglement between systems of qubits, is considered. The model of noise origination in polarizing qubits, which is determined by the presence of radiation spectral degree of freedom that appears during light propagation inside the anisotropic medium with dispersion, is studied. An approximate analytical model for calculation of the effect that phase plate has on a polarizing state with dispersion is suggested. The results, which are used to define a measure that characterizes the ability of a phase plate to preserve quantum polarizing state purity, are obtained.

## ACKNOWLEDGEMENTS

This work was partly supported by Russian Foundation of Basic Research (projects 13-07-00711, 14-01-00557), and by the Program of the Russian Academy of Sciences in fundamental research.

## REFERENCES

[1] Nielsen M. A., Chuang I. L. Quantum Computation and Quantum Information (Cambridge Series on Information and the Natural Sciences). Cambridge University Press, 2000. 675 p.
[2] Bogdanov Yu. I., Chernyavskiy A. Yu., Holevo A. S., Luckichev V. F., Nuyanzin S. A., Orlikovsky A. A. Mathematical modeling of quantum noise and the quality of hardware components of quantum computers, 2012, arXiv:1207.3313 [quant-ph].
[3] Abragam A., The Principles of Nuclear Magnetism. Oxford: Clarendon Press, 1961. 599 p.
[4] Holevo A. S., Quantum systems, channels, information. De Gruyter (2013).
[5] Bogdanov Yu. I., Chernyavskiy A. Yu., Holevo A. S., Lukichev V. F., Orlikovsky A. A., Bantysh B. I. Creating, maintaining, and breaking of quantum entanglement in quantum operations // Proceedings of SPIE. 2013. T. 8700. Art. 87001B.
[6] Bogdanov Yu. I., Bantysh B. I., Kalinkin A. A., Kulik S. P., Moreva E. V., Shershulin V. A.. Optical polarization echo: Manifestation and study by methods of quantum tomography of states and processes // Journal of Experimental and Theoretical Physics, 2014, V. 118, No. 6, P. 845-855.